\documentstyle[12pt]{article}
\begin{document}

\title{Universal effective action \\
for $O(n)$--symmetric $\lambda \phi^4$ model \\
from renormalization group}
\author{\large A.~I.~Sokolov,  E.~V.~Orlov,  V.~A.~Ul'kov,
and S.~S.~Kashtanov \\[3mm]
\em Department of Physical Electronics and Department of Physics, \\
\em Saint Petersburg Electrotechnical University, \\
\em Professor Popov Street 5, St. Petersburg, 197376, Russia}
\date{July 13, 1998}
\maketitle
\begin{abstract}
The RG expansions for renormalized coupling constants $g_6$ and $g_8$ of the
3D $n$-vector model are calculated in the 4-loop and 3-loop
approximations respectively. Resummation of the RG series for $g_6$
by the Pad$\acute e$-Borel-Leroy technique results in
the estimates for its universal critical values $g_6^*(n)$ which are arqued
to deviate from the exact numbers by less than 0.3 $\%$. The RG expansion
for $g_8$ demonstrates stronger divergence being much less 
suitable for getting proper numerical estimates.
\end{abstract}

Recently, the free energy (effective action) and, in particular,
higher-order renormalized coupling constants $g_{2k}$ for the basic
models of phase transitions became the target of intensive theoretical
studies (see, e. g. \cite{S98,PV} and references therein).
These constants enter the scaling equation of state and thus play a key
role at criticality. Along with critical exponents, they are universal,
i.e. possess, under $T \to T_c$, numerical values which depend only on
the space dimensionality and the symmetry of the order parameter.
Calculation of the universal critical values of $g_6$, $g_8$, etc.
for the 3D Ising model by different methods showed that the field-
theoretical RG approach in fixed dimensions yields the most accurate
numerical estimates. It is a consequence of a rapid convergence of the
iteration schemes originating from RG expansions \cite{S98,SOU,GZ}.
It is natural, therefore, to use the field
theory for calculation of renormalized higher-order coupling constants
for the 3D $n$-vector model. In the report, the 3D RG expansion for
the renormalized coupling constants $g_6$ and $g_8$ will be found and 
the universal critical values of $g_6$ will be estimated.

The 3D $O(n)$-symmetric model is described at criticality by Euclidean
scalar field theory with the Hamiltonian
\begin{equation}
H =
\int d^3x \Biggl[{1 \over 2}( m_0^2 \varphi_{\alpha}^2
 + (\nabla \varphi_{\alpha})^2)
+ \lambda (\varphi_{\alpha}^2)^2 \Biggr] ,
\label{eq:1} \\
\end{equation}
where $m_0^2$ is proportional to
$T - T_c^{(0)}$, $T_c^{(0)}$ being the phase transition
temperature in the absence of the order parameter fluctuations.
The fluctuations give rise to many-point correlations
$<\varphi(x_1) \varphi(x_2)...\varphi(x_{2k})>$ and,
correspondingly, to higher-order terms in the expansion of the
free energy in powers of the magnetization $M$:
\begin{equation}
F(M, m) = F(0, m) + \sum_{k = 1}^{\infty} g_{2k} m^{3 - k(1 + \eta)} M^{2k},
\label{eq:2} \\
\end{equation}
where $m$ is a renormalized mass, $\eta$ is a Fisher exponent, and
$g_{2k}$ are dimensionless coupling constants. Let, as usually, $g_2 = 1/2$.
Then $g_4$, $g_6$, $g_8$,... will acquire, under $T \to T_c$, the universal
values.

The asymptotic critical values of $g_4$, $g_4^*(n)$, determining
critical exponents and other universal quantitites, have been found
from the 6-loop expansion for RG $\beta$-function \cite{S98,BNM,LGZ,AS};
they are known with rather high accuracy. On the contrary, the information
about the universal numbers $g_6^*$, $g_8^*$, etc. for $O(n)$-symmetric model
is very poor today. To estimate them, we calculate corresponding RG series
and perform their resummation by means of the Pade-Borel-Leroy technique.
The RG series for $g_6$ and $g_8$ are obtained from
conventional Feynman graph expansions for the 6-point and 8-point
vertices in terms of the bare coupling constant $\lambda$. In its turn,
$\lambda$ is expressed perturbatively via the renormalized coupling
constant $g_4$. Substituting then the series for $\lambda$ into the
"bare" expansions, we obtain the RG expansions for $g_6$ and $g_8$.

As was shown \cite{SOU,SUO}, the 1-, 2-, 3-, and 4-loop contributions to
$g_6$ are formed by 1, 3, 16, and 94 one-particle irreducible Feynman
graphs, respectively. In this case, the calculations just described give:
\begin{eqnarray}
g_6 = {9 \over \pi} g_4^3 \Biggl[ {n + 26 \over {27}} -
{17~n + 226 \over {81 \pi}} g_4 + (0.000999164~n^2
+ 0.14768927~n + 1.24127452) g_4^2
\nonumber \\
-~(- 0.00000949~n^3 + 0.00783129~n^2 + 0.34565683~n
+ 2.14825455) g_4^3 \Biggr]. \qquad \qquad
\label{eq:3}
\end{eqnarray}

In the case of $g_8$, the 1-, 2-, and 3-loop contributions are
given by 1, 5, and 36 graphs respectively \cite{SUO}. Corresponding
RG expansion is found to be:
\begin{eqnarray}
g_8 = -{81 \over {2 \pi}} g_4^4 \Biggl[ {n + 80 \over {81}} -
{81~n^2 + 7114~n + 134960 \over {13122 \pi}} g_4
\qquad \qquad \qquad \qquad
\nonumber \\
+ (0.00943497~n^2 + 0.60941312~n + 7.15615323) g_4^2 \Biggr].
\qquad
\label{eq:4}
\end{eqnarray}

Being a field-theoretical perturbative expansions these series
are divergent (asymptotic).
To get reasonable numerical estimates for $g_6^*$ and $g_8^*$ some
procedure making them convergent should be applied.
The Borel-Leroy transformation
\begin{equation}
f(x) =  \sum_{i = 0}^{\infty} c_i x^i =
\int\limits_0^{\infty} t^b e^{-t} F(xt) dt , \qquad
F(y) = \sum_{i = 0}^{\infty} {c_i \over (i+b)!} y^i .
\label{eq:5}
\end{equation}
can play a role of such a procedure. Since the RG series considered
turns out to be alternating the analytical continuation of the Borel-Leroy
transform may be then performed by using Pad$\acute e$ approximants [L/M].

For $g_6$ we have the 4-loop RG expansion and can construct,
in principle, three different Pad$\acute e$ approximants: [2/1], [1/2],
and [0/3]. To obtain proper approximation schemes, however, only diagonal
[L/L] and near-diagonal Pad$\acute e$ approximants should be employed.
That's why further we limit ourselves with approximants [2/1] and [1/2].
Moreover, the diagonal Pad$\acute e$ approximant [1/1] will be also dealt
with although this corresponds to the usage of the lower-order, 3-loop
approximation.

The algorithm of estimating $g_6^*$ we use here is as follows.
Since the Taylor expansion for the effective action contains as
coefficients the ratios $R_{2k} = g_{2k}/g_4^{k-1}$
we work with the RG series for $R_6$. It is resummed in three different
ways based on the Pad$\acute e$ approximants just mentioned.
The Borel-Leroy integral is evaluated as a function of the parameter
$b$ under $g_4 = g_4^*(n)$. For the fixed point coordinate
$g_4^*(n) $ the values extracted from the six-loop RG expansion
are adopted \cite{S98,BNM}.
The optimal value of $b$ providing the fastest convergence of the
iteration scheme is then determined. It is deduced from the condition
that the Pad$\acute e$ approximants employed should give, for $b = b_{opt}$,
the values of $R_6^*$ which are as close as possible to each other.
Finally, the average over three estimates for $R_6^*$ is found and
claimed to be a numerical value of this universal ratio.

The results of our calculations are presented in Table 1. It contains
numerical estimates for $g_6^*$ resulting from the 4-loop RG expansion
(column 3) and their analogs given by the Pad$\acute e$-Borel resummed 3-loop
RG series \cite{S98} (column 4). As is seen, with increasing $n$ the
difference between the 4-loop and 3-loop estimates rapidly diminishes:
being small (0.9 \%) even for $n = 1$, it becomes negligible at $n = 10$
and practically disappears for $n \ge 14$. Such a behaviour is quite natural
since with increasing $n$ the approximating properties of RG series for
$g_6$ are improving \cite{S98}.

How close to the exact values of $g_6^*$ may the numbers in column 3 be?
To clear up this point, let us compare our 4-loop estimate for $R_6^*$
at $n = 1$ with those obtained recently by an analysis of the 5-loop
scaling equation of state for the 3D Ising model \cite{GZ,GZ98}. R. Guida
and J. Zinn-Justin have obtained $R_6^* = 1.644$ and, taking into
account some additional information, $R_6^* = 1.643$, while our estimate is
$R_6^* = 1.648$. Keeping in mind that the exact value of $R_6^*$ should lie
between the 4-loop and 5-loop estimates (the RG series is
alternating), our estimate can differ from the exact number by no more
than 0.3 \%. Since for $n > 1$ the RG expansion (\ref{eq:3}) should
provide better numerical estimates than in the Ising case, this
value (0.3 \%) represents an upper bound for the deviation of
the numbers in column 3 of Table 1 from their exact counterparts.

It is interesting to compare our estimates with those obtained
by other methods. Since 1994, the universal values of the sextic coupling
constant for the 3D $O(n)$--symmetric model were estimated by solving
the exact RG equations \cite{TW}, by lattice calculations \cite{R}, and by
a constrained analysis of the $\epsilon$--expansion \cite{PV}; corresponding
results are collected in columns 5, 6, and 7 of Table 1 respectively.
As is seen, they are, in general, in accord with ours. For large $n$, our 
estimates are consistent also with those given by the $1/n$-expansion
which are presented in column 8.

The RG expansion for the octic coupling constant $g_8$ turns out to be
much worse than the series (\ref{eq:3}) from the point of view of their
summability. Indeed, the series (\ref{eq:4}) diverges considerably stronger
and is one term shorter than that for $g_6$. It implies that the only
Pade approximant -- [1/1] -- may be really used in a course of the
resummation of this series. In the Ising case $n = 1$, such a simple
Pade-Borel procedure, when applied to the 3-loop RG expansion for $g_8$,
was found to lead to rather crude numerical estimates \cite{SUO}. 
As our analysis shows, with increasing $n$ the situation becomes better
but, nevertheless, the RG estimates for $g_8^*(n)$ remain much less 
accurate than those obtained for the sextic coupling constant. 
Corresponding numerical results will be published elsewhere.

\newpage
\begin{table}
\caption{Our estimates of universal critical values of the renormalized
sextic coupling constant for the 3D $n$-vector model (column 3).
The fixed point coordinates $g^*$ are taken from [5]
($1 \le n \le 3$) and [1] ($4 \le n \le 40$).
The $g_6^*$ estimates extracted from the Pade-Borel resummed
3-loop RG expansion (column 4), from the exact RG equations
(column 5), obtained by the lattice calculations (column 6),
resulting from a constrained analysis of the $\epsilon$-expansions
(column 7), and given by the $1/n$-expansion (column 8) are presented
for comparison.}

\label{table}
\begin{tabular}{c||c||c|c|c|c|c|c}
\hline
$n$ & $g^*$ & $g_6^*$ & $g_6^* \cite{S98}$ & $g_6^* \cite{TW}$ &
$g_6^* \cite{R}$ & $g_6^* \cite{PV}$ & $g_6^*~(1/n)$ \\
\hline
  & 2 & 3 & 4 & 5 & 6 & 7 & 8 \\
\hline
1 & 1.415 & 1.608 & 1.622 & 1.52 & 1.92(24) & 1.609(9) & \\
\hline
2 & 1.406 & 1.228 & 1.236 & 1.14 & 1.27(25) & 1.21(7) & \\
\hline
3 & 1.392 & 0.951 & 0.956 & 0.88 & 0.93(20) & 0.931(46) & \\
\hline
4 & 1.3745 & 0.747 & 0.751 & 0.68 & 0.62(15) & 0.725(29) & 1.6449 \\
\hline
5 & 1.3565 & 0.596 & 0.599 &  &   &  & 1.0528 \\
\hline
6 & 1.3385 & 0.483 & 0.485 &  &   &  & 0.7311 \\
\hline
7 & 1.321 & 0.396 & 0.398 &   &   &  & 0.5371 \\
\hline
8 & 1.3045 & 0.329 & 0.331 &   &   & 0.319(4) & 0.4112 \\
\hline
9 & 1.289 & 0.277 & 0.278 &   &   &  & 0.3249 \\
\hline
10 & 1.2745 & 0.235 & 0.236 &   &   &  & 0.2632 \\
\hline
12 & 1.2487 & 0.174 & 0.175 &   &   &  & 0.1828 \\
\hline
14 & 1.2266 & 0.134 & 0.134 &   &   &  & 0.1343 \\
\hline
16 & 1.2077 & 0.105 & 0.105 &   &   & 0.1032(4) & 0.1028 \\
\hline
18 & 1.1914 & 0.0845 & 0.0847 &   &   &  & 0.0812 \\
\hline
20 & 1.1773 & 0.0693 & 0.0694 &   &   &  & 0.0658 \\
\hline
24 & 1.1542 & 0.0487 & 0.0488 &   &   &  & 0.0457 \\
\hline
28 & 1.1361 & 0.0360 & 0.0361 &   &   &  & 0.0336 \\
\hline
32 & 1.1218 & 0.0276 & 0.0276 &   &   & 0.0275(1) & 0.0257 \\
\hline
36 & 1.1099 & 0.0218 & 0.0218 &   &   &  & 0.0203 \\
\hline
40 & 1.1003 & 0.0176 & 0.0176 &   &   &  & 0.0164 \\
\hline
\end{tabular}
\end{table}
\end{document}